# Implementations of quantum and classical gates with linear optical devices and photon number quantum non-demolition measurement for polarization encoded qubits

João Batista Rosa Silva and *Rubens Viana Ramos

*Department of Teleinformatic Engineering, Federal University of Ceara, Campus do Pici, 710, C.P. 6007, 60755-640, Fortaleza-Ceará, Brazil*

**Abstract**

Aiming the construction of quantum computers and quantum communication systems based on optical devices, in this work we present possible implementations of quantum and classical CNOTs gates, as well an optical setup for generation and distribution of bipartite entangled states, using linear optical devices and photon number quantum non-demolition measurement.

Quantum computation has attracted much attention since it was shown by Shor and Grover the possibility to implement quantum algorithms able to realize, respectively, factoring [1] and searching [2] in a much faster way than any other known classical algorithm. In fact, there is an enormous expectative that it will be possible to solve in a fast way several hard classical problems that are NP (non polynomial) when treated with classical computation. However, in despite of its potentialities, to build a quantum computer is a hard task. In fact, the most fundamental task, to build reliable quantum gates, is still a challenge. Up to now, several different technologies have been tested in the construction of quantum gates. Between them, the most important and promising are optical and photonic devices [3-7], quantum dots [8], superconducting devices [9,10], semiconductors [11,12] and nuclear magnetic resonance [13-15]. Each one of these technologies has its own advantages and disadvantages and, up to now, it is not clear which of them will dominate in the near future. However, optical and photonic devices technology has attracted much attention because, among other reasons, light polarization is a qubit relatively easy to create, to process and to detect. In fact, the first quantum teleportation protocol, that is a quantum computation primitive, was implemented in an optical experiment using light polarization [16]. Further, it is useful to remind that any single-qubit gate for polarization encoded qubit can be easily constructed using a polarization rotator between two retarder plates. On the other hand, it is well known that any quantum circuit (for qubits processing) can be built using only single-qubit gates and CNOTs [17]. Hence, in order to build a quantum computer using optical and photonic devices one must be able to construct reliable CNOT gates. Nonetheless, the construction of such gate requires highly nonlinear optical materials and such materials are currently not available. In order to circumvent this problem, some implementations of CNOTs using linear optics have been proposed. The price to be paid is the non-deterministic character of the produced gates [3,4,6]. That is, sometimes the CNOT operation fails. On the other hand, for quantum communication purposes, the creation and distribution of entanglement is an important task. Sharing a pair of maximally entangled photons, two users can implement quantum communication protocols or quantum non-local operations as a non-local CNOT. In this direction, the aim of this work is to present possible implementations of classical and quantum CNOT gates and an optical setup for entanglement creation and distribution, for polarization encoded qubits, using linear optical devices and photon number quantum non-demolition (QND) measurement.

___________
*Corresponding author
*E-mail addresses*: joaobrs@deti.ufc.br, rubens@deti.ufc.br

The QND measurement is a powerful resource of quantum computation [18] that induces the nonlinearity necessary for CNOT implementation [19]. Theoretical details of QND measurements can be found in [20,21]. Here, our interest is to use QND measurement to detect the presence of single-photons. Such task was nicely discussed in [22,23] and here we present a brief review of the main ideas. Basically, the photon number QND measurement can be achieved using a cross-Kerr nonlinearity, whose Hamiltonian is of the form $H_{QND} = \hbar \kappa a_s^+ a_s a_p^+ a_p$, where $\kappa$ is the strength of the nonlinearity, $a_s^+(a_s)$ is the creation (annihilation) operator of the signal mode while $a_p^+(a_p)$ is the creation (annihilation) operator of the probe mode. One can check that

$$e^{i\frac{H_{QND}t}{\hbar}}\left[a|0\rangle_s + b|n\rangle_s\right]|\alpha\rangle_p = a|0\rangle_s|\alpha\rangle_p + b|n\rangle_s|\alpha e^{in\kappa t}\rangle_p. \tag{1}$$

where $t$ is the time of interaction. Hence, the number state remains unaffected by the interaction while the coherent state $|\alpha\rangle$ picks up a phase shift directly proportional to the number of photons in the number state. The scheme for single-photon QND measurement is shown in Fig. 1.

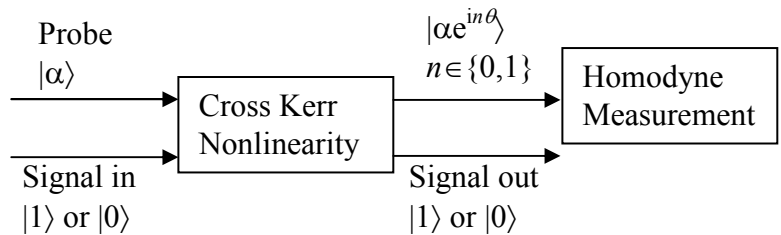

Fig. 1 – Single-photon quantum non-demolition measurement.

The phase measurement of the probe beam, via homodyne detection, determines the number of photons of the signal mode. For this, the Kerr medium is placed inside a Mach-Zehnder interferometer, as shown in Fig. 2.

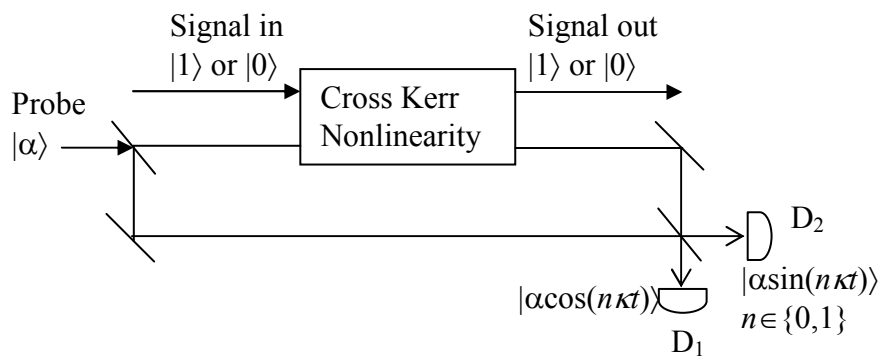

Fig. 2 – Phase measurement via homodyne detection.

The interferometer is tuned in such a way that a click in $D_1$ means no photon and a click in $D_2$ means one photon. In order to be able to detect the presence of a single-photon the condition $\kappa t > \pi/\left(2\sqrt{\langle n \rangle}\right)$ must be satisfied [24, 25], where $\langle n \rangle$ is the mean photon number of the coherent state. Hence, the coherent state amplitude must be large enough to obey the condition $|\alpha|\kappa t \gg 1$. This condition is not easy to obtain with the present tiny nonlinearity of optical fiber, for example, but the use of giant Kerr nonlinearity achievable with electromagnetically induced transparency [26] can make the single-photon QND measurement a reality in the near future.

An interesting and useful optical setup for QND polarization preserving single-photon detection, proposed in [27], is shown in Fig. 3.

The functioning of setup shown in Fig. 3 is straightforward, if there is a phase shift of $\theta$ in the coherent state, then a photon was present, otherwise there was no photon. Since the presence of a phase shift in the coherent state does not identify the polarization (vertical or horizontal) the superposition is preserved.

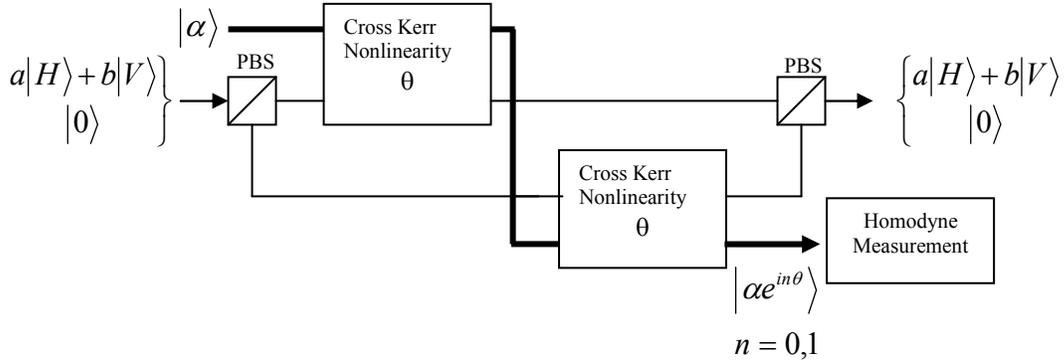

Fig. 3 – QND polarization preserving single-photon detection.

Hereafter we will show how to implement, for polarization encoded qubit, using linear optics and single-photon QND detection, CNOT gates and a setup for entanglement creation and distribution.

The optical setup presented in Fig. 4 implements a deterministic classical-CNOT without using single-photon QND measurement.

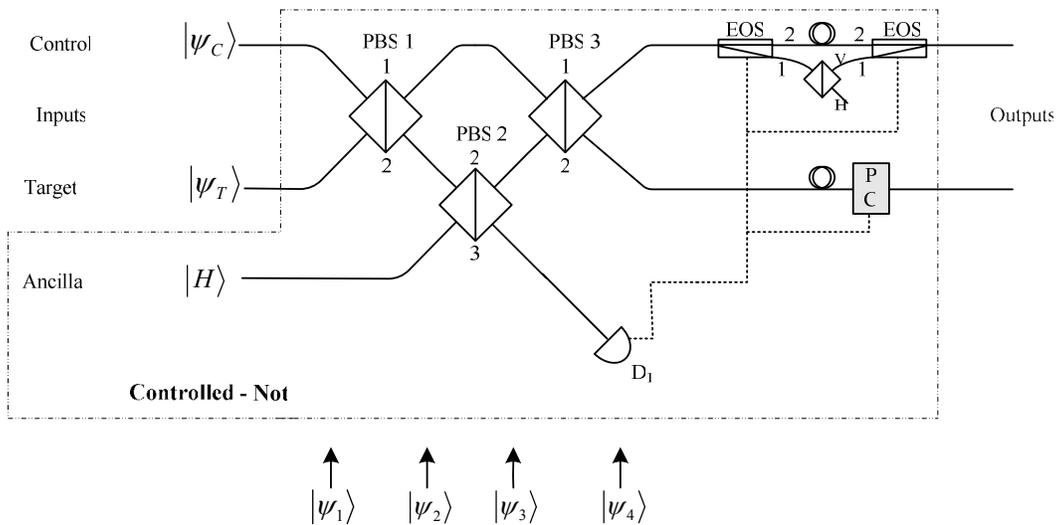

Fig. 4 - Deterministic classical-CNOT (without single-photon QND measurement). PBS – Polarization beam splitter. PC – Pockels cell. $D_1$ – Single-photon detector. EOS – Electro-optic switch.

The functioning of the setup shown in Fig. 4 is quite simple. Firstly, the polarization beam splitters (PBS) transmit vertical polarization and reflect horizontal polarization. When the input (control and target qubits) is the state $|HH\rangle$ or $|HV\rangle$, there will be detection in $D_1$ and nothing has to be done. On the other hand, when the input is the state $|VH\rangle$ or $|VV\rangle$ there will be no detection in $D_1$, and two photons with different polarization will appear at the upper arm. In this case, the electro-optic switches (EOS) and the Pockels cell (PC) are turned on. The PC will invert the target qubit, as is required for a CNOT operation, and the setup formed by two EOS and one PBS takes the horizontally polarized photon from the upper arm. Hence, the setup of Fig. 4 implements the true table $\{|HH\rangle \to |HH\rangle,\ |HV\rangle \to |HV\rangle,\ |VH\rangle \to |VV\rangle,\ |VV\rangle \to |VH\rangle\}$. This classical-CNOT needs one ancilla that is consumed in half of the cases.

Now we will show an implementation of a deterministic classical-CNOT using single-photon QND measurement and without ancilla. The setup proposed can be seen in Fig. 5.

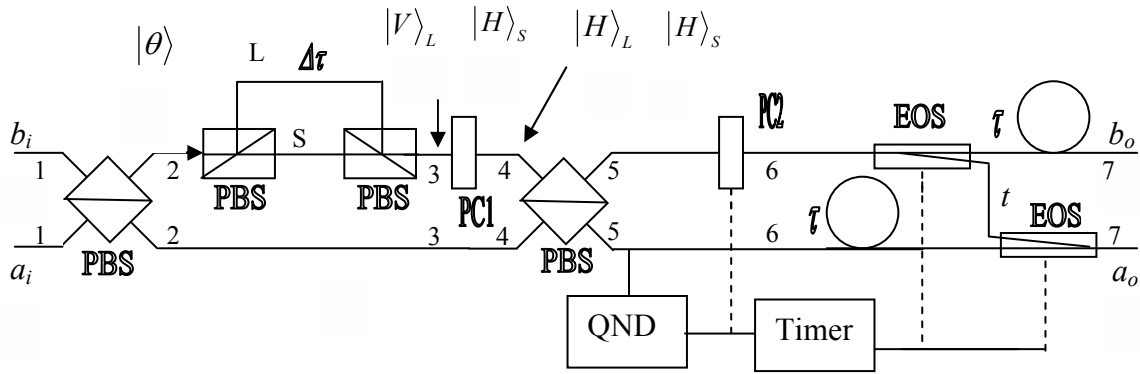

Fig. 5 – Classical-CNOT gate implemented with linear optical devices and photon QND measurement. $\tau$, $\Delta\tau$ and $t$ are delays obeying the relation $\tau = \Delta\tau + t$. the numbers 1, 2, …, 7 are specific places where the total state is analyzed.

It works as follow: The input modes are $a_i$ and $b_i$ while the output modes are $a_o$ and $b_o$. When mode $b_i$ ($a_i$) is a photon having horizontal polarization, it leaves the PBS (polarization beam splitter) at the upper (lower) arm. When mode $b_i$ ($a_i$) is a photon having vertical polarization, it leaves the PBS at the lower (upper) arm. At the upper arm, a pulse having horizontal (vertical) polarization takes the shorter (longer) path. Between the third (from left to right) PBS and the Pockels cell PC1 there will be a time-bin qubit. The Pockels cell PC1 is activated in order to give a rotation of $\pi/2$ ($|H\rangle \to |V\rangle$, $|V\rangle \to |H\rangle$) only in the polarization of the delayed pulse. The pulses arriving at the fourth PBS are guided, according to their polarization, in the same way as happen in the first PBS. The QND measurement checks if there is a photon in the lower arm of the fourth PBS. If there is not a photon there, a signal is sent to rotate the polarization of $\pi/2$, via Pockels cell PC2 in the upper arm, and both EOS are activated in order to guide only the delayed pulse from the upper arm to the lower arm. Since $\tau = \Delta\tau + t$ both pulses will arrive at the output at the same time. The synchronization of the two Pockels cells and the two electro-optic switches is an important point in the gate implementation. For this task a timer solves the problem. The following table shows the operation when the inputs are states of the canonical basis $\{|H\rangle,|V\rangle\}$.

| 1($b_i$) | 1($a_i$) | 3($u$) | 3($l$) | 5($u$) | 5($l$) | 6($u$) | 6($l$) | 7 ($b_o$) | 7 ($a_o$) |
|---|---|---|---|---|---|---|---|---|---|
| $|H\rangle$ | $|H\rangle$ | $|H\rangle_S$ | $|H\rangle_S$ | $|H\rangle_S$ | $|H\rangle_S$ | $|H\rangle_S$ | $|H\rangle_S$ | $|H\rangle$ | $|H\rangle$ |
| $|H\rangle$ | $|V\rangle$ | $|V\rangle_L|H\rangle_S$ | $|0\rangle$ | $|H\rangle_L|H\rangle_S$ | $|0\rangle$ | $|V\rangle_L|V\rangle_S$ | $|0\rangle$ | $|V\rangle$ | $|V\rangle$ |
| $|V\rangle$ | $|H\rangle$ | $|0\rangle$ | $|H\rangle_S|V\rangle_S$ | $|V\rangle_S$ | $|H\rangle_S$ | $|V\rangle_S$ | $|H\rangle_S$ | $|V\rangle$ | $|H\rangle$ |
| $|V\rangle$ | $|V\rangle$ | $|V\rangle_L$ | $|V\rangle_S$ | $|H\rangle_L|V\rangle_S$ | $|0\rangle$ | $|H\rangle_S|V\rangle_L$ | $|0\rangle$ | $|H\rangle$ | $|V\rangle$ |

Table 1 – Operation of the classical-CNOT gate shown in Fig. 5. *u* means upper arm and *l* means lower arm and the numbers 1, 3, 5, 6 and 7 are specific locals at the setup, as shown in Fig. 5.

From table 1 one easily realizes that mode $a_i$ is the control bit, while $b_i$ is the target bit. The performance of the classical-CNOT gate presented depends strongly on the performance of the QND measurement. If it is deterministic, then the classical-CNOT will also be.

In order to implement a quantum CNOT a number of solutions have been proposed [3,6,27,28,29]. Here, we will discuss the usefulness of the solution presented in [27,28]. The setup proposed in those references, is shown in Fig. 6.

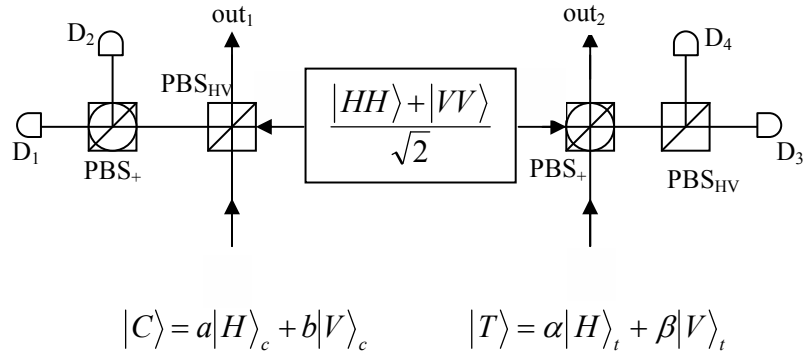

Fig. 6 – CNOT gate implemented with polarization beam splitters (PBS) and two-photon entangled state. PBS$_{HV}$: PBS in horizontal-vertical basis. PBS$_+$: PBS in diagonal basis ($\pi/4$, $3\pi/4$). D$_{1-4}$ are single-photon detectors.

The optical setup shown in Fig. 6 implements the CNOT gate only when a single-photon is detected in D$_1$ or D$_2$ and another single-photon is detected in D$_3$ or D$_4$. The output state of the setup in Fig. 6 is:

$$|\Psi_f\rangle = \frac{1}{4}\left[|+,H\rangle|\psi_1\rangle + |-,H\rangle|\psi_2\rangle + |+,V\rangle|\psi_3\rangle + |-,V\rangle|\psi_4\rangle\right] + \frac{\sqrt{3}}{2}|\Psi_u\rangle \qquad (2)$$

$$|\psi_1\rangle = \left[a\alpha|HH\rangle_{ct} + a\beta|HV\rangle_{ct} + b\alpha|VV\rangle_{ct} + b\beta|VH\rangle_{ct}\right] \qquad (3)$$

$$|\psi_2\rangle = \left[-a\alpha|HH\rangle_{ct} - a\beta|HV\rangle_{ct} + b\alpha|VV\rangle_{ct} + b\beta|VH\rangle_{ct}\right] = [(XZX)\otimes I]|\psi_1\rangle \qquad (4)$$

$$|\psi_3\rangle = \left[a\alpha|HV\rangle_{ct} + a\beta|HH\rangle_{ct} + b\alpha|VH\rangle_{ct} + b\beta|VV\rangle_{ct}\right] = [I\otimes X]|\psi_1\rangle \qquad (5)$$

$$|\psi_4\rangle = \left[-a\alpha|HV\rangle_{ct} - a\beta|HH\rangle_{ct} + b\alpha|VH\rangle_{ct} + b\beta|VV\rangle_{ct}\right] = [(XZX)\otimes I][I\otimes X]|\psi_1\rangle \qquad (6)$$

In (2) $|\Psi_u\rangle$ is the useless part that contains the situations where none or two photons were detected in $D_{1\text{-}2}$ and/or $D_{3\text{-}4}$. Further, $|+,H(V)\rangle$ means a single-photon going to $D_1$ and another single-photon going to $D_3$ ($D_4$), while $|-,H(V)\rangle$ means a single-photon going to $D_2$ and another single-photon going to $D_3$ ($D_4$). Observing (2) and (3) one sees that the success probability of CNOT operation is 1/16. However, if one uses single-qubit operations to correct the output state according to where the detections were obtained (detections in $D_2$ and $D_3$ → $XZX$ in the control qubit, detections in $D_1$ and $D_4$ → $X$ in the target qubit and detections in $D_2$ and $D_4$ → $XZX$ in the control qubit and $X$ in the target qubit) the probability of success goes to 1/4.

A crucial component in the CNOT implementation of Fig. 6 is the entangled pair of photons. If instead of $(|HH\rangle+|VV\rangle)/2^{1/2}$ the state $(|HV\rangle+|VH\rangle)/2^{1/2}$ is used, one would get the unitary operation $(X\otimes I)U_{CNOT}(X\otimes I)$, that is, a CNOT that inverts the target when the control qubit is $|H\rangle$. On the other hand, if one uses the disentangled state $(|HH\rangle+|VH\rangle)/2^{1/2}$ instead of the Bell state, the setup in Fig. 6 implements the identity operation. The main problems with CNOT implementation of Fig. 6 are its probabilistic behavior and the necessity of reliable entangled two-photon sources. However, it has been proposed a near determinist CNOT using quantum non-demolition measurement in [22].

Having the setup of Fig. 6 as inspiration, a simplified, and less efficient, CNOT, using only optical fiber devices, can be seen in Fig. 7. In this setup, the entangled pair is probabilistically (50%) produced by the first $PBS_{HV}$ having at its input the total state $[(|H\rangle+|V\rangle)/2^{1/2}] \otimes [(|H\rangle+|V\rangle)/2^{1/2}]$. In this case, the first $PBS_{HV}$ total output state is $[(|HH\rangle+|VV\rangle)/2^{1/2}]/2^{1/2}+[(|0,HV\rangle+|HV,0\rangle)/2^{1/2}]/2^{1/2}$. The optical mirror configuration makes necessary only one $PBS_{HV}$ and one $PBS_+$, instead of two of each of them, as happens in Fig. 6. On the other hand, four optical circulators are necessaries. The functioning of the CNOT shown in Fig. 7 is equal to CNOT of Fig. 6, but the former has half of the efficiency of the last, because of the probabilistic Bell state generation.

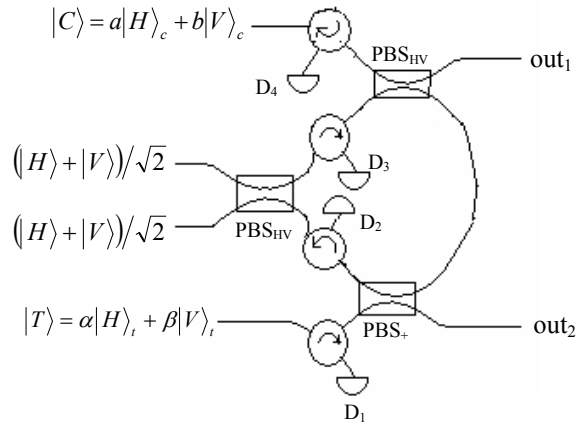

Fig. 7 – CNOT implementation using only optical fiber devices.

At last, let us now consider the optical setup shown in Fig. 8.

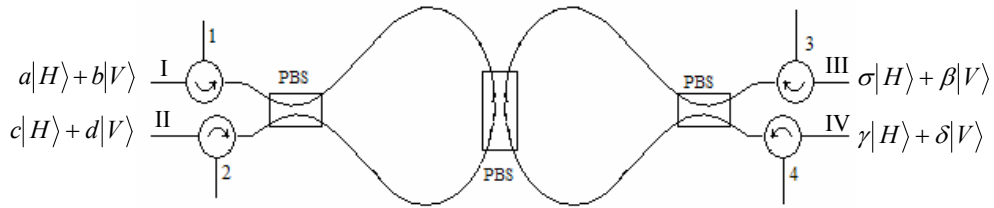

Fig. 8 – Optical setup for entangled bipartite state generation and distribution.

After some calculations one easily finds the output state

$$|\Psi\rangle_{1234} = ac\sigma\gamma|HHHH\rangle + ad\sigma\delta|VHVH\rangle + bc\beta\gamma|HVHV\rangle + bd\beta\delta|VVVV\rangle + |\Omega\rangle \quad (7)$$

where $|\Omega\rangle$ is the state containing terms with at least one output with zero photons. Using the QND polarization preserving photon detection of Fig. 3 in each output, one can be sure when a valid output state was obtained (one photon at each output). For the optical setup in Fig. 8, if we consider the qubits I and IV as input and qubits II and III as ancillas (both in the state $(|H\rangle+|V\rangle)/2^{1/2}$), the optical setup realizes, when the system does not fail, the transformation

$$(a|H\rangle+b|V\rangle)_I (\gamma|H\rangle+\delta|V\rangle)_{IV} \left(\frac{|H\rangle+|V\rangle}{2}\right)_{II} \left(\frac{|H\rangle+|V\rangle}{2}\right)_{III} \rightarrow (\gamma|HH\rangle+\delta|VV\rangle)_{13}(a|HH\rangle+b|VV\rangle)_{24} \quad (8)$$

Hence, the setup of Fig. 8 can be used to generate two bipartite states with arbitrary (and controlled) amount of entanglement. For example, if one uses a not gate (a halfwave plate) at the output 3 and $a=b=\gamma=\delta=2^{-1/2}$, then the output state is $(|HV\rangle+|VH\rangle)/2^{1/2})_{13}(|HH\rangle+|VV\rangle)/2^{1/2})_{24}$, that could be used as resource for two CNOT gates of the type shown in Fig. 6, one activated by $|V\rangle$ and other activated by $|H\rangle$. Setup of Fig. 8 can also be used for quantum communication purpose. Suppose qubits I and II belong to Alice and qubits III and IV belong to Bob (the central PBS can be in Alice's or Bob's place). If once more Alice and Bob choose the parameters' values $a=b=\gamma=\delta=2^{-1/2}$, then the total output is $(|HH\rangle+|VV\rangle)/2^{1/2})_{13}(|HH\rangle+|VV\rangle)/2^{1/2})_{24}$ meaning that Alice and Bob share two maximally entangled state that they can use for quantum key distribution or quantum teleportation, for example. Further, sharing a Bell state, Alice and Bob can implement a non-local CNOT of the type shown in Fig. 6, with control qubit belonging to Alice and target qubit belonging to Bob. Obviously, for a reliable use of setup shown in Fig. 8 as entanglement generator and distributor, the polarization states must be transmitted in an undisturbed way through the long optical fiber. This can be achieved, obeying the necessaries restrictions, with the optical setups proposed in [30].

In summary, we have presented two optical setups for deterministic classical-CNOTs construction, one with ancilla and without single-photon QND detection, and the other without ancilla and using single-photon QND detection. Following, after discussing the implementation of a probabilistic quantum CNOT using a Bell state as resource, we presented an optical setup for quantum CNOT gate implementation using only simple optical devices. It has low efficiency (the probability of success is 1/32) but it is very simple to construct. At last, we presented an optical setup for entanglement generation and distribution. Basically, two distant parts send two qubits and receive two qubits. The received qubits are entangled in such way that the parts share two pairs of entangled states. The amount of entanglement of the shared pairs is controlled by the input qubits parameters. Once both users share an entangled state they can run, for example, quantum key distribution or quantum teleportation protocols, or to realize a non-local CNOT operation using the CNOT implementation that requires a Bell state.

**Acknowledgements**

This work was supported by the Brazilian agency FUNCAP.